\documentstyle[twocolumn,aps,prb,graphics]{revtex}

\begin{document}
\twocolumn[\hsize\textwidth\columnwidth\hsize\csname@twocolumnfalse\endcsname
\author{A. A. Teklu,$^{1}$ R. G. Goodrich,$^{1}$ N. Harrison,$^{2}$ D.
Hall,$^{3}$
\and Z. Fisk,$^{3}$ and D. Young$^{3}$ \and $^{1}$ {\small Department of
Physics and Astronomy, Louisiana State University, Baton Rouge, LA 70803.}
\and $^{2}$ {\small National High Magnetic Field Laboratory, LANL, MS-E536,
Los Alamos, NM 87545.} \and $^{3}${\small \ National High Magnetic Field
Laboratory, Florida State University, Tallahassee, FL 32310.}}
\title{Fermi Surface Properties of Low Concentration Ce$_{x}$La$_{1-x}$B$_{6}$:
dHvA.}
\date{March 20, 2000.}
\maketitle

\begin{abstract}
The de Haas-van Alphen effect is used to study angular dependent extremal
areas of the Fermi Surfaces (FS) and effective masses of Ce$_{x}$La$_{1-x}$B$%
_{6}$ alloys for $x$ between 0 and 0.05. The FS of these alloys was
previously observed to be spin polarized at low Ce concentration ( $x$ =
0.05). This work gives the details of the initial development of the
topology and spin polarization of the FS from that of unpolarized metallic
LaB$_{6}$ to that of spin polarized heavy Fermion CeB$_{6}$ .

\end{abstract}
]

\section{Introduction}

The rare earth (RE) and divalent hexaborides have a variety of electrical,
magnetic and thermodynamic properties and all have the same cubic structure.
Among these materials are metallic LaB$_{6}$ [1], Kondo insulating SmB$_{6}$
[2], semi-metallic CaB$_{6}$[3 - 5], heavy Fermion (HF) CeB$_{6}$[6], and
ferromagnetic EuB$_{6}$ [7]. Extensive experimental and theoretical
investigations have been done in order to understand their varying physical
properties. One of the most decisive techniques to study the electronic
properties of these materials is the de Haas - van Alphen (dHvA) effect with
which the extremal cross-sectional areas of the Fermi surface (FS) and
effective masses can be measured accurately. Pure LaB$_{6}$ and pure CeB$_{6}
$ have been studied using this technique, having nearly identical prolate
ellipsoidal FS's, with the FS of CeB$_{6}$ being larger than that of LaB$_{6}
$ by about 10\% [8,9]. For example, the values of the dHvA frequencies for
LaB$_{6}$ and CeB$_{6}$ for the same minimum FS ellipsoid cross-section are
7.89kT and 8.66kT respectively. Yet, the effective masses are quite
different being 0.65m$_{e}$ and 30m$_{e}$(at 5-7 T) for LaB$_{6}$ and CeB$%
_{6}$ respectively[8,9], where m$_{e}$ is the free electron mass.

There have been several electrical, magnetic and thermal studies carried out
to explore how this transition from light metallic LaB$_{6}$ to the HF CeB$%
_{6}$ takes place when La ions are gradually replaced by Ce ions that
introduce 4f electrons into the metal. In addition to this, experimental
work has been carried out, using the dHvA effect at high magnetic fields (%
$>$20 T), to explore the development of the HF behavior in Ce$%
_{x} $La$_{1-x}$B$_{6}$ [10]. Here, it was reported that both the FS
topology and effective masses transform continuously from that of pure LaB$%
_{6}$ to that of pure CeB$_{6}$ as the Ce concentration $x$ is increased
from $0$ to $1$. Furthermore, beginning at very low values of $x$ (about
0.05), the contribution to the dHvA signal was observed to originate from
only a single spin FS sheet.

Here we report detailed dHvA measurements, using both the field modulation
technique at intermediate fields (6-15 T) and cantilever torque measurements
to 30 T to investigate how the spin polarization manifests itself in the
topological changes of the FS, and changes in effective masses of Ce$_{x}$La$%
_{1-x}$B$_{6}$ alloys for 0 $\leq $ $x$ $\leq $ 0.05. The results of these
measurements are then compared with the previous pulsed field measurements
[10] and found to be in excellent agreement. In this paper, the spin
polarization of the FS\ is investigated both qualitatively and
quantitatively, and it is found that the spin up component dominates the
dHvA signal as the Ce concentration increases.

\section{Experiment and Sample Preparation}

Single crystals of Ce$_{x}$La$_{1-x}$B$_{6}$ with $x$ = 0, 0.01, 0.02, 0.03,
0.04, and 0.05 were grown in Al flux in the shape of rectangular
parallelepipeds (1$\times $0.5$\times $2 mm) with each face along a [100]
axis of the cubic structure [11]. Most of the dHvA magnetization
measurements on these samples were made in a 0$-$18 T superconducting magnet
using the field modulation technique where a small time oscillating field $%
h_{0}$ is superimposed on a steady state field. The sample was placed inside
an astatic pair of pick up coils that was balanced to 1 part in 10$^{4}$ at
zero field. The remaining imbalance signal voltage in the pick up pair due
to the magnetization of the sample oscillating with changing magnetic field
was detected using a lock-in amplifier.

Using field modulation, constant angle dHvA measurements were made in the
field range 6-15 T with the field sweeping at a rate of 0.05 T/min and the
field direction rotated within (100) crystal plane. From these dHvA
oscillations one can determine the extremal cross-sectional areas and
effective masses from the temperature dependence of signal amplitudes over
the entire FS. The sample was further rotated continuously in a fixed field
of 10 T to observe the detailed dependences of the FS cross sectional areas
on angle. For these measurements, the sample rotator angle was calibrated so
that the orientation of the sample was known to a precision of rotation of
0.1${{}^{\circ }}$. The above measurements were made at five or six
different temperatures depending on the sample in the temperature range of
1.4 K to 4.2 K with the sample immersed in a pumped He$^{4}$ bath. The
sample temperature was measured using a calibrated Cernox thermometer and
the vapor pressure of the bath. The field was calibrated with NMR.

To verify the reproducibility of these results, torque measurements on an $x$%
= 0.01 Ce sample were made to 30 T at the National High Magnetic Field
Laboratory, Tallahassee, FL.

\section{Experimental Results}

Figure 1 shows typical dHvA oscillations for Ce$_{x}$La$_{1-x}$B$_{6}$ ($x$=
0.01) for $\theta $ = 0${{}^{\circ }}$ (i.e., the [100] direction) in the
field range of 10$-$11 T and at a temperature of 1.4 K. A discrete Fourier
transform (DFT) of the signal is shown on the same graph. For measurements
on pure LaB$_{6}$, the frequency of the minimum area or $\alpha _{3}$ orbit
is found to be 7.894 $\pm $ 0.004 kT, which is in good agreement with the
original measurements of Arko et al.[1]. With this as a point of reference,
complete angular dependent studies of the frequencies in all of the Ce
concentrations were made. This study was made in order to check the
assumption that the FS is represented by an ellipsoid of revolution, because
this assumption had previously been used to calculate FS volumes [10]. Both
constant angle field sweeps and constant field angle sweeps were obtained.
Example data from the constant field rotation measurements for $x$ = 0.01 at
T = 1.4 K and H = 10 T is shown in the inset of Figure 2. The oscillations
with angle are caused by the fact that the dHvA phase, $2\pi F(\theta )/H,$
changes by $2\pi $ for each complete oscillation as $\theta $ is varied. The
angular variation of $F$ can be determined from these rotation measurements
using the counting method first implemented by Halse [12]. Using this
technique, the angular variations of the minimum area or $\alpha _{3}$ and
maximum area or $\alpha _{12}$ orbits were obtained. As a further check,
field dependent measurements and DFTs were also made at several angles. An
example of the complete data for $x$ = 0.01 is shown in Figure 2.

The effective masses of the different samples (0 $\leq $ $x$ $\leq $ 0.05)
are extracted from the temperature dependences of the oscillation
amplitudes. In comparison, the value of the effective mass of the $\alpha
_{3}$ orbit for x = 0 or pure LaB$_{6}$ was found to be 0.66$\pm $0.03m$_{e}$
compared to the results of Arko et al. [1] which give 0.65m$_{e}$. Thus, the
two values are the same to within experimental uncertainty.

\section{Discussion}

It is well known that the cross-sectional area $A$ of the FS in atomic units
(a.u.) is related to the dHvA frequency $F$ by the Onsager relation [13],

\begin{equation}
A(a.u)=(\frac{2\pi e}{\hbar c})F=2.673\times 10^{-5}F
\end{equation}
where $F$ is the dHvA frequency in tesla. From the measured values of the
dHvA frequencies, the extremal cross - sectional areas of the $\alpha _{3}$
and $\alpha _{1,2}$ orbits for the field applied along the [100] crystal
axis of Ce concentrations between $x$ = 0 and $x$ = 0.05 can be calculated.
As shown in Figure 3, both of the frequencies corresponding to the $\alpha
_{3}$ and $\alpha _{1,2}$ orbits are observed to increase with $x$. From
these two frequencies, the volume of the FS, and hence the number of charge
carriers per unit volume can be evaluated assuming that the FS is an
ellipsoid of revolution. The $x$ dependence of the carrier density, n,
calculated in this manner is shown in the inset of Figure 3 [10].

It has been reported that both LaB$_{6}$ and CeB$_{6}$ have similar prolate
electron ellipsoidal FS's situated at the six X points of the cubic
Brillouin zone(BZ) that overlap along the $\Gamma $R symmetry axes [10,14].
This situation is shown schematically in Figure 4. The minimum
cross-sectional area of the ellipse corresponding to the $\alpha _{3}$-orbit
can be measured directly by applying the magnetic field \textbf{H} along the
[100] axis, while the maximum area for the $\alpha _{1,2}$ orbit is only
observed through magnetic breakdown (MB) for this same field orientation
[10]. MB through the necks also leads to a multitude of frequencies $\alpha
_{1,2}+n\rho $ within the $\Gamma XM$ plane, $\rho $ being the orbit
associated with a small FS orbit inside the necks and n is an integer [15].
The value of $\rho $ is $\sim $ 4 - 8 T [16], which is smaller than our
experimental uncertainty, and therefore cannot be resolved at the fields
used for our measurements.

The expected angular dependence of the dHvA frequency from an ellipsoid of
revolution is [17]:

\begin{equation}
\frac{1}{F(\theta )}=A\cos ^{2}\theta +B\sin ^{2}\theta +C\sin \theta \cos
\theta
\end{equation}
Here $F$($\theta )$ is the dHvA frequency and $\theta $ is the angle of
rotation from the principal axis of the ellipsoid normal to the field
direction. A fit of the data to this equation is a useful criterion for
deciding whether or not the FS is an ellipsoid of revolution [17]\textit{. }
We are interested here in the angular variations of the $\alpha _{3}$ and $%
\alpha _{1,2}$ orbits. If the [100] axis is rotated through an angle
relative to the field direction in the (100) plane, the cross-sectional area
of the FS normal to the field direction corresponding to the semi-minor axis
of the ellipse, or the $\alpha _{3}$ orbit, increases while the one
corresponding to the semi-major axis or the $\alpha _{1,2}$ orbit decreases,
and these two areas or frequencies become degenerate at 45${{}^{\circ }}$
(see Figure 2). These two frequencies and their fit to Equation 2 are
plotted on the same graph in Figure 2, showing excellent agreement between
the expected angular dependence from ellipsoidal FS and the data. This same
agreement is obtained for all of the samples with 0 $\leq $ $x$ $\leq $ 0.05
measured here. Thus, all of the measurements support the assumption used
previously [9,10] that the FS is an ellipsoid of revolution.

Figure 5 shows the concentration ($x$) dependence of m* for the $\alpha _{3}$
orbit and the curve fit to the data has the quadratic form m$%
_{e}(c+bx+ax^{2})$. According to Gor'kov and Kim, a linear dependence of the
specific heat coefficient $\gamma $ (proportional to m*) and the magnetic
susceptibility $\chi $ of Ce and U based alloys would be a signature of
contributions from independent impurity centers [18]. However, at larger
values of concentration in a system of localized spins, the linear
dependence on $x$ does not hold and Gor'kov and Kim [18] calculated an
additional $x^{2}$ correction term to both the specific heat and magnetic
susceptibility, using the Fermi liquid formulation. Therefore, the very good
fit of our data to a quadratic equation relating m* and $x$ would indicate
that impurity centers are coupled even at low Ce concentrations. This
observation is consistent with the model of coupled Ce atoms giving rise to
the antiquadrupolar state in CeB$_{6}$ [19]. We would like to point out that
the effective mass m* could be spin-dependent. At this stage in the
analysis, the effective mass m* is the one determined from the temperature
dependence of the dHvA signal which has contributions from spin up and spin
down electrons. The spin dependence of m* will be discussed later in Section
V.

In the usual case when both spin states have the same mass, the magnetic
field dependence of the dHvA amplitude can be used to determine the average
Dingle temperature $\overline{T}_{D}$ for the two spin states (see Section
V). The effect of finite relaxation time due to impurity or point defects is
to broaden the Landau levels leading to a reduction in amplitude roughly
equivalent to that which would be caused by a rise of temperature to $%
\overline{T}_{D}.$ The field dependence of the dHvA amplitude can be
expressed as [17]

\begin{equation}
A_{p}=\frac{C_{p}TH^{-n}R_{D}}{\sinh (\alpha pT/H)}
\end{equation}
where $A_{p}$ is the amplitude of the $p^{th}$ harmonic, $R_{D}$ $%
=exp(-\alpha p\overline{T}_{D}/H)$ is the Dingle reduction factor, $p$ is
the harmonic number, $\alpha $ =$14.69(m*/m_{e})T$ $/K$, and $C_{p}$ and $n$
depend on the particular method of measurement. For the torque method the
value of $n=-1/2$ and a plot of $ln(A_{p}H^{-1/2}\sinh (\alpha pT/H))$
versus $1/H$ yields a straight line with a slope of $\alpha p\overline{T}%
_{D} $ and a linear fit to the data gives $m^{*}\overline{T}_{D}$. A Dingle
plot for the high field cantilever data for 1\%Ce in LaB$_{6}$ at T=1.73 K
and in the field range $10-25$ T is given in Figure 6. From the slope of the
straight line the value of the average Dingle temperature $\overline{T}_{D}$
was found to be 3.5 K at high fields on the assumption that m = 0.73m$_{e}$
for both spin states. We have analyzed the field dependence of the amplitude
for all six samples (0 $\leq $ $x$ $\leq $ 0.05) using field modulation in
the range 7 $\leq $ B $\leq $ 15 T and we found that $m^{*}\overline{T}_{D}$
is the same within the measurement uncertainty. This means that the
arbitrary substitution of La by Ce (or Ce by La) contributes little or
nothing to the mean free path \textsl{l} of the dominant spin channel which
is given by

\begin{equation}
\frac{l}{l_{c}}=\omega _{c}\tau
\end{equation}
where \textsl{l}$_{c}$ is the cyclotron length for a particular orbit given
by [10]

\begin{equation}
l_{c}=\left( \frac{2\hbar F}{eB^{2}}\right) ^{1/2}.
\end{equation}
The average scattering rate is related to the average Dingle temperature by
the relation

\begin{equation}
\overline{T}_{D}=\frac{\hbar }{2\pi k\tau }.
\end{equation}
Thus, since $m^{*}\overline{T}_{D}$ is independent of $x$, the mean free
path \textit{l} is also independent of $x$ for a given field range. This
observation is in agreement with the results of Goodrich et. al. [10] that
the dominant source of scattering should originate from other forms of
crystallographic imperfections not from the Ce or La impurities.

\section{Spin Dependent Scattering (SDS)}

One of the effects of an applied magnetic field is to lift the spin
degeneracy of the energy levels and the contributions to the dHvA signal
from the spin-up and spin-down electrons. In conventional metals, the effect
of the Zeeman splitting is to reduce the amplitude by a spin reduction
factor $R_{S}$ given by[17]

\begin{equation}
R_{S}=\cos (\frac{1}{2}p\pi g\frac{m^{*}}{m_{e}})\equiv \cos (p\pi S)
\end{equation}
where $g$ is the spin-splitting factor and m$_{e}$ is the free electron
mass. We had reported earlier that above 5\% Ce in LaB$_{6}$, the
contribution to the dHvA amplitude originates from a single spin FS. One
explanation of this observation is that scattering from spin fluctuations
does not occur with equal strength for the two spin directions. There is a
large negative magnetoresistance [20] in Ce$_{x}$La$_{1-x}$B$_{6}$ alloys
that can be explained by the suppression of spin fluctuation scattering.
However, from magnetoresistance measurements one cannot determine if one or
two spin states are contributing to the scattering. Part of the purpose of
the present work was to study in detail how the single spin dHvA signal
develops in CeB$_{6}$ from the two spin signal in pure LaB$_{6}$.

In the presence of high magnetic fields, if there is only one spin
contribution to the signal, then a plot of $\ln (A_{p}/$ $p^{1/2})$ against
the harmonic number, $p,$ yields a straight line because the spin splitting
reduction factor as is given by Equation 7 is no longer present. However, if
there are contributions to the dHvA amplitude from spin-up and spin-down
states, there is spin reduction of the amplitude and we observe non linear
dependence of $\ln (A_{p}/$ $p^{1/2})$ on $p$ as shown in Figure 7 for 0 $%
\leq $ x $\leq $ 0.05 except for $x$ $\geq $ 0.05, the concentration at
which only one spin component is observed.

For pure LaB$_{6}$, the dHvA amplitudes associated with the spin-up and
spin-down electrons are equal. However, if magnetic impurities are involved,
we could expect spin dependent scattering and the amplitudes for spin-up and
spin-down oscillations could be unequal corresponding to unequal Dingle
temperatures. So, the signal amplitude measured with the field modulation
technique, which has contributions from both the spin-up and spin-down
components, has to be modified in order to account for differences in Dingle
temperatures and masses between the two spin channels. In the first harmonic
detection, this signal voltage is related to the oscillatory magnetization
by [21]

\begin{equation}
\widetilde{V}(\zeta )=G\sum_{p}\widetilde{M}_{p}J_{1}(p\Lambda )\sin (p\zeta
+\theta _{p}),
\end{equation}
where $G$ represents the system gain, $\widetilde{M}_{p}$ is the
magnetization due to the $p^{th}$ harmonic of the dHvA signal, $J_{1}$ is a
Bessel function of order one, $\Lambda $ = $2\pi h/H^{2},$ $h$ is the
modulation amplitude, $\zeta $ $=2\pi F/H,$ $\sigma $ $=$ $\pm 1,$ and $%
\theta _{p}$ is the phase. The magnetization $\widetilde{M}$ can be written
[21]:

\begin{equation}
\widetilde{M}=\sum\limits_{p=1}^{\infty }\sum\limits_{\sigma
}C_{p}D^{p}E^{\sigma p}\sin (p\zeta +p\frac{\pi }{4}-\sigma p\pi S)
\end{equation}
where

\begin{eqnarray}
D &=&\exp (-Km^{*}\overline{T}_{D}/H), \\
\overline{T}_{D} &=&(T_{D}^{\downarrow }+T_{D}^{\uparrow })/2 \\
E &=&\exp (-Km^{*}(\delta T_{D})/H) \\
\delta T_{D} &=&(T_{D}^{\downarrow }-T_{D}^{\uparrow })/2
\end{eqnarray}
and

\begin{equation}
C_{p}=\frac{\nu TF}{(A^{"}p\hbar )^{1/2}}\frac{1}{\sinh (pKm^{*}T/H)}
\end{equation}
where $\nu $ $=1.304\times 10^{-5}(Oe^{1/2}/K),$ $K$ $=$ $14.69(T/K),$ $F$
is the dHvA frequency and $A$ is the extremal cross sectional area of the FS.

If the phase difference $\phi _{p}(=p\pi S)$ between spin-up and spin-down
oscillations is field dependent, the first two harmonics of the
magnetization $\widetilde{M}$ may also be written as

\begin{eqnarray}
\widetilde{M}_{1} &=&C_{1}\left[ z\sin (\psi +\frac{\phi }{2})+z^{\prime
}\sin (\psi -\frac{\phi }{2})\right] \\
&=&C_{1}(z^{2}+z^{\prime 2}+2zz^{\prime }\cos \phi )^{1/2}\sin (\psi +\theta
_{1})
\end{eqnarray}
and

\begin{equation}
\widetilde{M}_{2}=C_{2}(z^{4}+z^{\prime 4}+2z^{2}z^{\prime 2}\cos (2\phi
))^{1/2}\sin (2\psi \mp \frac{\pi }{4}+\theta _{2})
\end{equation}
where $z$ is the Dingle reduction factor for the spin up electrons, $%
z^{\prime }$ is the Dingle reduction factor for the spin down electrons, and
$\psi $ is defined to be $(2\pi F/H)\pm \pi /4$ where the upper sign is for
a minimum FS area and the lower is for a maximum FS area. Other higher
harmonics can be written in a similar way. The relative phase between the
spin up and down components of the signal is given by

\begin{equation}
\tan \theta p=\frac{z^{p}-z^{\prime p}}{z^{p}+z^{\prime p}}\tan (\frac{p\phi
}{2})
\end{equation}
and the spin up and the spin down Dingle reduction factors for the $p^{th}$
harmonic are given by

\begin{equation}
z^{p}=\exp (-p\alpha T_{D}^{\uparrow }/H)
\end{equation}
and

\begin{equation}
z^{\prime p}=\exp (-p\alpha T_{D}^{\downarrow }/H)
\end{equation}

The relative phases are obtained by fitting the data to the first three
harmonics of the magnetization. From the measured signal harmonic amplitude
ratios $M_{2}/M_{1},$ $M_{3}/M_{1}$ and the relative phases between the
harmonics ($\theta _{2}-$ $2\theta _{1})$ and ($\theta _{3}-3\theta _{1})$ ,
we calculate, at a given H and T, the values of the amplitudes $D$ and $E$
(i.e., $m^{*}$ $\overline{T_{D}}$ and $m^{*}\delta T_{D})$ , and the value
of $S$. Once the value of $S$ is known, we can determine the amplitude ratio
$z^{\prime }/z$ (or $(m^{\downarrow }T_{D}^{\downarrow }-m^{\uparrow
}T_{D}^{\uparrow })$ ). The average Dingle temperature, $\overline{T_{D}},$
is spin independent while the difference $\delta T_{D}$ = $(T_{D}^{\uparrow
}-T_{D}^{\downarrow })$ $/2$ shows the spin-dependence. As mentioned
earlier, the effective mass m$^{*}$ could also depend on spin. Therefore,
since in all of the above expressions the product $m^{*}T_{D}$ where both m*
and T$_{D}$ are spin-dependent always occurs, it is not trivial to single
out the spin dependence of either one. Therefore, we will first assume that $%
m^{\uparrow }=m^{\downarrow }$ so that $C_{p}$ $($ Eqn. 14) is the same for
both spin states. The concentration dependence of $m(T_{D}^{\downarrow }$ $-$
$T_{D}^{\uparrow }$ $)$ is then calculated from the amplitude ratios $%
M_{2}/M_{1},$ $M_{3}/M_{1}$ and relative phases. Next, we assume $%
m^{\uparrow }\neq m^{\downarrow }such$ that $C_{p}^{\downarrow }\neq $ $%
C_{p}^{\uparrow }$ and again calculate the x dependence, now of $%
(m^{\downarrow }T_{D}^{\downarrow }-m^{\uparrow }T_{D}^{\uparrow })$. Figure
8 shows the concentration dependence of both $m(T_{D}^{\downarrow }$ $-$ $%
T_{D}^{\uparrow }$ $)$ and $(m^{\downarrow }T_{D}^{\downarrow }-m^{\uparrow
}T_{D}^{\uparrow })$ and there is clear evidence of SDS. In other words, if
there is no SDS, then $\delta T_{D}$ $=$ $T_{D}^{\downarrow }-$ $%
T_{D}^{\uparrow }$ $=$ $0$, that is the slope of the $x$ dependence of $%
(m^{\downarrow }T_{D}^{\downarrow }-m^{\uparrow }T_{D}^{\uparrow })$ is
non-zero. The circles represent the case that $m^{\uparrow }=m^{\downarrow },
$ while the squares represent the case $m^{\uparrow }\neq m^{\downarrow }.$
The slope of the line corresponding to $m^{\uparrow }\neq m^{\downarrow }$
is approximately eight times that corresponding to $m^{\uparrow
}=m^{\downarrow }$ indicating that $m^{\downarrow }T_{D}^{\downarrow }$
becomes greater than $m^{\uparrow }T_{D}^{\uparrow }$ as the Ce
concentration increases, that additional increase in slope arising from the
difference in mass.

For LaB$_{6}$, the two spin components have equal amplitudes and the
amplitude ratio $z^{\prime }/z$ is equal to one or $\delta T_{D}$ $=$ $
T_{D}^{\downarrow }-$ $T_{D}^{\uparrow }$ $=$ $0$. In other words, the
scattering rates for spin up and spin down are equal. As the Ce
concentration is increased to 5\%, the ratio of the spin-up to the spin-down
amplitude increases confirming the complete observed spin polarization of
the FS at 5\%Ce in LaB$_{6}$ to one spin channel which is the spin up.

There are two contrasting theories concerning whether the FS polarizes to
the spin up or spin down state. The first is the theory developed by
Wasserman et al. [22] for quantum oscillations in heavy fermion materials.
This model, along with its zero field predecessor [23], is successful in
accounting for the heavy effective masses as well as small topological
changes in the FS caused by the presence of additional f electrons. However,
this model also predicts that the dHvA signal is dominated by the spin down
channel, and that its associated effective mass should decrease in a
magnetic field. While apparent evidence for this was reported in very heavy
compounds such as CeCu$_{6}$ [24], it is CeB$_{6}$ that shows perhaps the
most dramatic mass changes with increasing magnetic field [25], but the
polarity of the spin was not identified.

A large number of measurements have been performed on CeB$_{6}$ [26-34],
which is regarded as a typical dense Kondo lattice with a very low Kondo
temperature of $1-2$ K. Previous experimental data have appeared to be
entirely consistent with the theoretical model of Wasserman et al. [22],
that is, the effective mass is dramatically suppressed in a magnetic field
[25]. Recent measurements [9] have shown that in addition to the suppression
of the effective mass, there is notable deformation in the topology of the
FS in a magnetic field, and this result is not entirely consistent with the
mean field theory of Ref. [22]. One aspect of the result that does appear to
be consistent with this theoretical model, though, is that the dHvA signal
originates from only a single spin FS sheet [9], even though the theory must
be fundamentally incorrect because it predicts the wrong spin state to be
observed.

All of the dHvA measurements on CeB$_{6}$ are made in the high magnetic
field regime well above the metamagnetic transition where the dipole moments
of the f electrons are essentially aligned. According to Edwards and Green
[35], in this regime the theory developed by Wasserman et al. [22] is no
longer applicable. This is due to the fact that this is a mean field
approach in which the interactions are assumed not to change in a magnetic
field. Making such a description of the dHvA effect in HF systems is really
only valid at low magnetic fields, that is, magnetic fields less than the
Kondo temperature scale. Edwards and Green [35] instead make the analogy of
a HF compound in a magnetic field to a itinerant ferromagnet, in which spin
fluctuations play a decisive role. Edwards and Green [35] also anticipate
that the dHvA effect should be dominated by only a single spin, but the up
spin instead of the down spin. Therefore, one can see that our measurements
are in agreement with the predictions of Edwards and Green that the down
spin mass enhancement is larger than that of the up spin and does not
contribute to the dHvA signal amplitude.

As further verification of this mass difference, we use Equations (19) and
(20) to write

\begin{equation}
-\frac{H}{K}\ln (z/z^{\prime })=(m^{\downarrow }-m^{\uparrow
})T+(m^{\downarrow }T_{D}^{\downarrow }-m^{\uparrow }T_{D}^{\uparrow })
\end{equation}
The quantity on the left hand side of Equation 21 is calculated for each
value of $x$ and plotted versus T in Figure 9. It can be seen that the data
is linear in $T$ with a slope of $(m^{\downarrow }-m^{\uparrow })$ and
intercept $(m^{\downarrow }T_{D}^{\downarrow }-m^{\uparrow }T_{D}^{\uparrow
})$. The value of $(m^{\downarrow }-m^{\uparrow })$ ranges from 0.003 for $x$
= 0 or pure LaB$_{6}$ to 0.09 for $x$ = 0.05 respectively. In addition, the
value of $(m^{\downarrow }T_{D}^{\downarrow }-m^{\uparrow }T_{D}^{\uparrow
}) $ ranges from 0.03 for $x$ = 0 to 0.3 for $x$ = 0.05. Thus, for pure LaB$%
_{6} $ , which is not spin polarized, $m^{\downarrow }=$ $m^{\uparrow }$ as
expected. As the Ce concentration increases to 5\%, $(m^{\downarrow }-$ $%
m^{\uparrow })$ increases with $m^{\downarrow }$ being greater than $%
m^{\uparrow }$ by about 10\%. If this mass difference continues to increase
with $x$ the observed discrepancy between specific heat and dHvA mass
measurements in CeB$_{6}$ is explained. Moreover, the difference in the
scattering parameter,$(m^{\downarrow }T_{D}^{\downarrow }-m^{\uparrow
}T_{D}^{\uparrow })$ , increases with the Ce concentration confirming that
the observed FS\ is due only to the spin up state. These observations lead
us to the conclusion that it is the combination of $\Delta m^{*}$ and SDS
that takes the down spin out of the dHvA signal. Therefore, both $\Delta
m^{*}$ and SDS are equally important in understanding many of the properties
of CeB$_{6}$ .

\section{Conclusion}

We have performed a detailed microscopic dHvA study of low concentration (up
to x = 0.05) in Ce$_{x}$La$_{1-x}$B$_{6}$ alloys and determined the
development of the size and geometry of the FS from that of spin unpolarized
LaB$_{6}$ to the spin polarized Ce$_{x}$La$_{1-x}$B$_{6}$ $(x\geq $ $0.05)$
alloys. We have shown:

\begin{itemize}
\item  The spin up signal amplitude dominates the dHvA signal when the Ce
concentration is $\geq $ 5\%.

\item  We determined that the down spin mass is greater than that of the up
spin and the spin down contribution to the dHvA signal amplitude is small.

\item  The angular dependence of the dHvA extremal areas of the FS show that
the assumption that the FS is an ellipsoid of revolution is valid for all
concentrations measured.

\item  The dependence of the effective mass on concentration is in agreement
with the existing theories of magnetic impurity interactions [18].
\end{itemize}

Overall, this work is the most detailed study using dHvA of alloy systems
involving magnetic ions with concentrations greater than 1\% that has been
reported.

{\large Acknowledgment}

A portion of this work was performed at the National High Magnetic
Laboratory, which is supported by NSF Cooperative agreement No. DMR -
9527035 and by the State of Florida. Additional support from the NSF
(DMR9971348) is acknowledged by Z. Fisk.

{\large References}

1. A. P. J. Arko, et al., \textit{Phys. Rev.} \textbf{B} \textbf{13}, 5240

\quad (1976)

2. P. Nyhus, et al., \textit{Phys. Rev.} \textbf{B} \textbf{55}, 12488-12496

\quad (1997)

3. H. C. Longuet-Higgins and M. de V. Roberts,

\quad \textit{Proc. Roy. Soc. London} \textbf{A\ 224}, 336-347 (1954)

4. W. A. C. Erkelens, et al., \textit{J. Magn. Magnet. }

\quad \textit{Mater.} \textbf{63/64}, 61-63 (1987)

5. D. P. Young, et al., \textit{Nature} \textbf{397}, 412(1999)

6. S. Sullow, et al., \textit{Phys. Rev.} \textbf{B} \textbf{57}, 5860-5869

\quad (1998)

7. L. Degiorgi, et al., \textit{Phys. Rev. Lett}. \textbf{79},

\quad 5134-5137 (1997)

8. A. P. J. van Deursen, Z. Fisk and A. R. de

\quad Vroomen, \textit{Solid State Commun}. \textbf{44}, 609 (1982)

9. N. Harrison, et al.,\textit{\ Phys. Rev. Lett}. \textbf{81}, 870

\quad (1998)

10. R. G. Goodrich, et al., \textit{Phys. Rev. Lett}., \textbf{82},

\quad 3669 (1999)

11. R. G. Goodrich, et al., \textit{Phys. Rev}. \textbf{B} \textbf{58}, 14896

\quad (1998)

12. M. R. Halse,\textit{\ Phil. Trans. Roy. Soc}. \textbf{A} \textbf{265},

\quad 53 (1969)

13. L. Onsager, \textit{Phil. Mag}. \textbf{43}, 1006 (1952)

14. Y. Onuki, et al., \textit{J. Phys. Soc. Jpn}. \textbf{58}, 3698

\quad (1989)

15. N. Harrison, et al.,\textit{\ Phys. Rev. Lett.}, \textbf{80}, 4498

\quad (1998)

16. Y. Ishizawa, et al., \textit{J. Phys. Soc. Jpn}, \textbf{48}, 1439

\quad (1980)

17. D. Shoenberg, \textit{Magnetic Oscillations in Metals}

\quad (Cambridge University Press, Cambridge, 1984)

18. L. P. K. Gor'kov and Ju H. Kim, \textit{Phys. Rev}.

\quad \textbf{B} \textbf{51}, 3970 (1995)

19. F. J. Ohkawa, J. \textit{Phys. Soc. Japan} \textbf{52}, 3897

\quad (1983)

20. N. Sato, et al., \textit{J. Phys. Soc. Jpn}. \textbf{54}, 1923

\quad (1985)

21. Alles, H. G., Higgins, R. J. and Lowndes,

\quad D. H. \textit{Phys. Rev. Lett.} \textbf{30}, 705 (1973)

22. A. Wasserman, M. Springford, and F. Han,

\quad \textit{J. Phys.:Condens. Matter }\textbf{3}, 5335 (1991)

23. J. W. Rasul, \textit{Phys. Rev}. \textbf{B} \textbf{39}, 663 (1989)

24. S. B. Chapman, et al., \textit{Physics Lett}. \textbf{B} \textbf{163},

\quad 361 (1990)

25. N. Harrison, et al., J. Phys.: \textit{Condens. Matter}

\quad \textbf{5}, 7435 (1993)

26. A. P. J. van Deursen et al., \textit{J. Less-Common}

\quad \textit{\ Mat}. \textbf{111}, 331 (1985)

27. W. Joss, et al., \textit{Phys. Rev. Lett}. \textbf{59}, 1609

\quad (1987)

28. R. Foro, et al.,\textit{\ J. Phys. Soc. Jpn} \textbf{57}, 2885

\quad (1988)

29. W. Joss, et al., \textit{de Physique} \textbf{49}, 747 (1988)

30. W. Joss, \textit{J. Mag. and Mag. Mat}.\textbf{84}, 264 (1990)

31. Y. Onuki, et al., \textit{Physica} \textbf{B} \textbf{163}, 100 (1990)

32. E.G. Haanappel, et al., \textit{Physica} \textbf{B177}, 181

\quad (1992)

33. H. Matsui, et al., \textit{Physica} \textbf{B} \textbf{186-188}, 126

\quad (1993)

34. R. Hill et al., \textit{Physica} \textbf{B} \textbf{230-232}, 114 (1997)

35. D. M. Edwards and A. C. Green, \textit{Z. Phys}. \textbf{B}

\quad \textbf{103}, 243 (1997).

\onecolumn
\widetext

\begin{center}
	\textbf{FIGURE CAPTIONS}
\end{center}

FIG. 1  An example of dHvA oscillations from the field modulation
measurements for $x$ = 0.01 sample for H along the [100] axis. The inset
shows the DFT of the oscillations for a field range of 10 $-$ 11 T.

FIG. 2 The dHvA frequencies F$_{3}$ and F$_{1,2}$ as a function of
orientation for $x$ = 0.01 from field sweep and rotation (high density
points in the Fig.) measurements. The solid lines are the fits to the to the
ellipsoid Eqn. 2. The inset shows the raw data from angular sweep
measurements at 10 T.

FIG. 3 Ce concentration dependence of the dHvA frequencies F$_{3}$ and F$%
_{1,2}$ at fields $10$ $-$ $11$ T. The inset shows the number of electrons
per unit volume, n, as a function of $x.$

FIG. 4 FS of LaB$_{6}$ .

FIG. 5 The Ce concentration dependence of cyclotron mass for the $\alpha
_{3} $ orbit at 10 T. The solid line is a quadratic fit to the data.

FIG. 6 Dingle plot for $x$ = 0.01 at 1.73 K in the field range of 10 $-$ 25
T using the cantilever technique.

FIG. 7 A plot of $\ln (A_{p}$ $/$ $p^{1/2})$ versus the harmonic number $p$
for $x$ = 0 to 0.05. Note that it becomes linear at $x$ = 0.05.

FIG. 8 Ce concentration dependence of $m^{\downarrow }T_{D}^{\downarrow }$ $%
- $ $m^{\uparrow }T_{D}^{\uparrow }$ .

FIG. 9 Temperature dependence of ($H/K)\ln (z/z^{\prime })$ for all the
alloys including pure LaB$_{6}.$ From the linear fits to the data, mass
differences between the two spin states were determined.

\end{document}